\documentclass[conference,10pt]{IEEEtran}
\IEEEoverridecommandlockouts
\usepackage{cite}
\usepackage{amsmath,amssymb,amsfonts}
\usepackage{algorithmic}
\usepackage{graphicx}
\usepackage{textcomp}
\usepackage{xcolor}
\usepackage{algorithm}
\usepackage[belowskip=-10pt]{caption}
\usepackage{caption}
\usepackage{subcaption}
\usepackage{booktabs}
\usepackage{array}
\usepackage{multirow}
\usepackage{xcolor}
\usepackage{soul} 
\usepackage[super]{nth}
\def\BibTeX{{\rm B\kern-.05em{\sc i\kern-.025em b}\kern-.08em
    T\kern-.1667em\lower.7ex\hbox{E}\kern-.125emX}}

\usepackage[top=0.75in,bottom=1in,left=0.625in,right=0.625in]{geometry}

\newcommand{\tsd}{360$^\circ$ }

\begin{document}

\title{ASL360: AI-Enabled Adaptive Streaming of Layered 360$^\circ$ Video over UAV-assisted Wireless Networks
%{\footnotesize \textsuperscript{*}Note: Sub-titles are not captured in Xplore and should not be used}
}
\author{
\IEEEauthorblockN{Alireza Mohammadhosseini\IEEEauthorrefmark{1}, Jacob Chakareski\IEEEauthorrefmark{1}, Nicholas Mastronarde\IEEEauthorrefmark{2}} 
\IEEEauthorblockA{\IEEEauthorrefmark{1}College of Computing, New Jersey Institute of Technology; \IEEEauthorrefmark{2}Dept. of Electrical Engineering, University of Buffalo}\thanks{This work has been supported in part by NSF awards CNS-2032033, 

CNS-2106150 and CNS-2346528, and in part by NIH award R01EY030470}
\vspace{-0.8cm}}

\maketitle

\begin{abstract} 
We propose \emph{ASL360}, an adaptive deep reinforcement learning-based scheduler for on-demand \tsd video streaming to mobile VR users in next generation wireless networks. We aim to maximize the overall Quality of Experience (QoE) of the users served over a UAV-assisted 5G wireless network. Our system model comprises a macro base station (MBS) and a UAV-mounted base station which both deploy mm-Wave transmission to the users. The \tsd video is encoded into dependent layers and segmented tiles, allowing a user to schedule downloads of each layer's segments. Furthermore, each user utilizes multiple buffers to store the corresponding video layer's segments. We model the scheduling decision as a Constrained Markov Decision Process (CMDP), where the agent selects Base or Enhancement layers to maximize the QoE and use a policy gradient-based method (PPO) to find the optimal policy. Additionally, we implement a dynamic adjustment mechanism for cost components, allowing the system to adaptively balance and prioritize the video quality, buffer occupancy, and quality change based on real-time network and streaming session conditions. We demonstrate that \emph{ASL360} significantly improves the QoE, achieving approximately 2 dB higher average video quality, 80\% lower average rebuffering time, and 57\% lower video quality variation, relative to competitive baseline methods. Our results show the effectiveness of our layered and adaptive approach in enhancing the QoE in immersive video streaming applications, particularly in dynamic and challenging network environments. %For future research, we plan to extend \emph{ASL360} by integrating computational resource allocation and evaluating its scalability in extended scenarios.
%This and the IEEEtran.cls file define the components of your paper [title, text, heads, etc.]. *CRITICAL: Do Not Use Symbols, Special Characters, Footnotes, 
%or Math in Paper Title or Abstract.
\end{abstract}

%\begin{IEEEkeywords}
%360 Video streaming, DASH streaming, Adaptive Viewport, Tiled streaming, Quality Emphasized Region, ROI Encoding
%\end{IEEEkeywords}

\section{Introduction}

Recent advances in wireless networks and virtual reality (VR) have significantly increased the demand for efficient \tsd video streaming \cite{Elbamby2018,Jacob2022}. These services introduce unique challenges in resource allocation and network management. Cellular networks often struggle to meet these stringent requirements, particularly under dynamic network conditions, which motivates the integration of Unmanned Aerial Vehicles (UAVs) to augment existing network infrastructure with flexible, high-capacity communications. UAV-assisted heterogeneous networks have emerged as a promising solution for supporting bandwidth-intensive services. These systems combine conventional macro base stations (MBSs) with UAV-mounted base stations employing mm-Wave communication links to enhance data rates and optimize the scheduling algorithm, thus significantly improving the VR users (VUs) QoE \cite{gupta2023,liu2018}.

To optimize the \tsd video streaming experience, various techniques have been proposed\cite{corbillon2017, hosseini2016}. For instance, viewport-dependent streaming methods such as tiled streaming based on Dynamic Adaptive Streaming over HTTP (DASH), divide panoramic video into spatial and temporal tiles\cite{Concolato2017,jeroen2019,Khodam2024}. Moreover, video content can be divided into scalable layers that can be efficiently and separately scheduled to increase the flexibility and performance of the system. The \tsd frames are encoded as the Base Layer (BL) which does not depend on any other layer and provides a basic quality, and the Enhancement Layers (ELs) which reference the BL, providing improved video quality. Prior studies \cite{hsu2020, bura2024} have investigated methods of optimizing video quality, considering practical factors like rebuffering or stall events.
However, existing solutions often overlook the dynamics of wireless communication and \tsd video streaming requirements. Consequently, data driven approaches like Deep Reinforcement Learning (DRL) have shown significant promise in addressing such dynamic optimization problems due to their ability to learn optimal policies directly from complex environments without explicitly modeling the environmental dynamics (see, e.g., \cite{Nuowen2019}, \cite{zhang2019}).
\begin{figure}[t]
\centering
\includegraphics[width=0.4\textwidth]{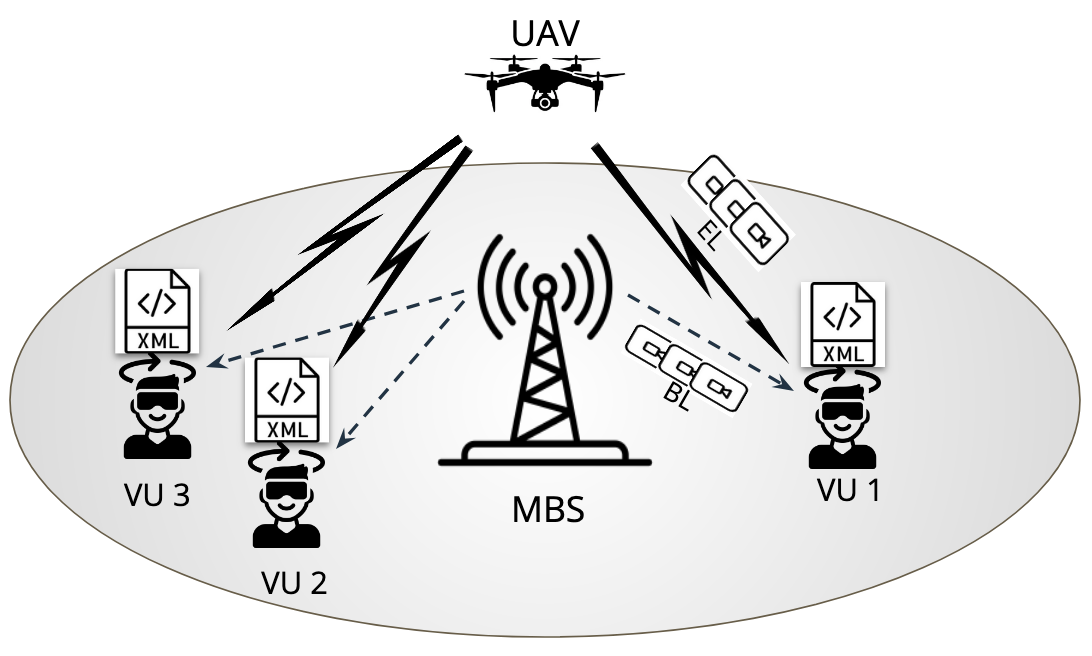}
\caption{Our MPEG-DASH based \tsd video streaming system.}
\label{sys}
\end{figure}
Motivated by these, we propose a DRL-based scheduling method specifically tailored for UAV-assisted wireless networks delivering layered \tsd video. Our proposed approach dynamically adjusts its decision-making strategy based on real-time network conditions, buffer status, and video-related QoE metrics. The contributions of our work includes layered \tsd video streaming system modeling and analysis, dynamic adaptation of reward weights to capture video streaming specific requirements over time, comprehensive modeling of user-centric QoE, and an evaluation of our approach against established baselines.
\footnote{This paper has been accepted for presentation at IEEE GLOBECOM 2025.}

%\section{Related Work} 
%Several approaches to model QoE investigated in the literature. In [?], the authors compared the required data rate for each service along with the achievable data model to model the QoE of the VU. In [?], the authors considered more practical aspects of QoE, including perceived bitrate and rebuffering events.

\section{System Model}
In this section, we introduce a UAV-assisted wireless network to serve layered \tsd video to VUs. The network comprises a macro base station (MBS) and a UAV-mounted base station. Let $\mathcal{B}$ denote the set of base stations and $\mathcal{V}$ represent the set of VUs. In the following sections, we describe each component of the system to lay the groundwork for the problem formulation.  %Since the system serves a single VU, we do not consider shared resources or multi-user interference.
%Rather than analytically modeling path loss or \textit{SNR}, we rely on empirical $5G$ throughput traces for both the MBS and the UAV connections and assume that the UAV's position is fixed throughout the streaming session (so we do not consider the UAVs trajectory). This approach supports realistic system modeling and allows us to focus on scheduling and QoE aspects. 

\subsection{\tsd Video Streaming Model}
To achieve high-quality \tsd video streaming, we consider an on-demand receiver-driven scalable video transmission system. In this system, the tiled DASH streaming approach is used to spatially and temporally segment the \tsd videos. Each panoramic video frame is divided into $\mathcal{L_X} \times\mathcal{L_Y}$ grid tiles. For the BL, all tiles within a Group of Pictures (GOP) are encoded at a lower quality providing smoother video playback. The bitrate of the BL GOP is obtained by summing bitrates of all tiles within that GOP. The tiles within the field of view (FoV) are encoded at a higher quality as the EL segment. The bitrate of the EL segment is determined by summing over all tiles’ bitrates in the user’s viewport. Using this methodology makes our approach unique to \tsd video streaming. The DASH server stores all BL and EL video segments and VUs request them from the appropriate base station according to the Media Presentation Description (MPD), as in Figure~\ref{sys}.

\subsection{Buffer Model}
\begin{figure}[h]
\centering
\includegraphics[width=0.4\textwidth]{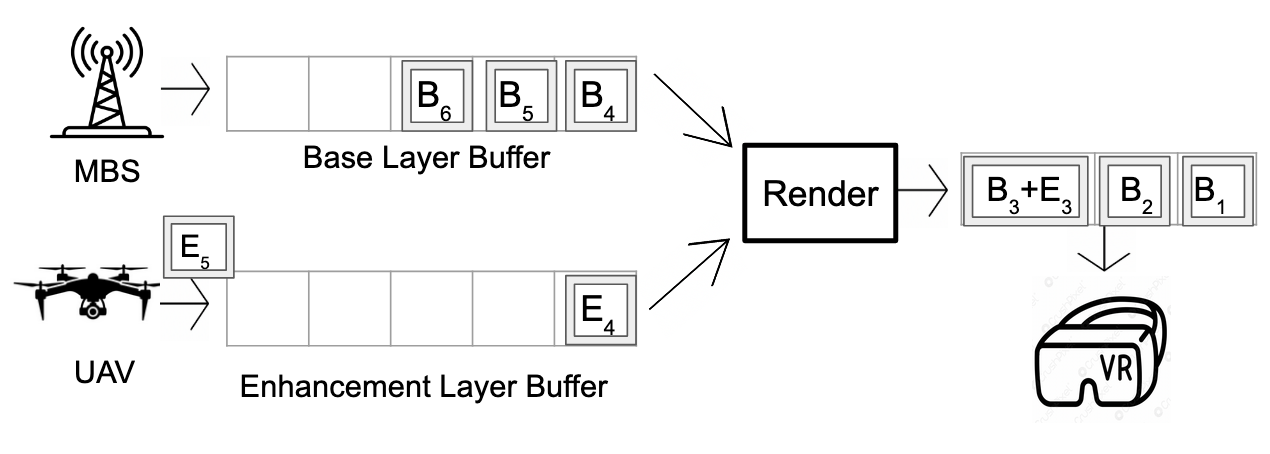}
\caption{Base and Enhancement Layer Buffer Models.}
\label{buffer}
\end{figure}

The receiver's buffer plays an important role in the on-demand streaming system, as the user's experience can be significantly affected by rebuffering events caused by buffer underflow. Also, it is important to maintain a balance between preloading video chunks to keep the buffers full and adapting to viewport changes in \tsd video streaming. A highly aggressive look-ahead strategy may lead to inefficiencies if the viewport prediction does not align with the user's actual head movement. Therefore, buffer management is crucial to maintain an optimal balance between rebuffering and responsiveness. In our system, the user utilizes two buffers, as illustrated in Figure~\ref{buffer}. These are: (i) Base Layer Buffer: Stores BL segments received from the MBS, and (ii) Enhancement Layer Buffer: Stores EL segments received from the UAV.
%\jc{You can bring the two bullet points inline to save space. No need to have them.}
% \begin{itemize}
%     \item The received BL segments from the MBS are stored in the BL Buffer.
%     \item The received EL segments from the UAV are stored in the EL Buffer.
% \end{itemize}

If a specific EL segment is available, it is merged with the corresponding BL segment, rendered, and then prepared for playback.
The buffer trade-off in our system suggests that a shorter EL buffer allows faster adaptation to viewport changes, ensuring that the user receives the most relevant content. On the other hand, a longer BL buffer provides smoother playback, mitigating potential interruptions due to network fluctuations but lower perceived quality.
 The client's buffers transition from one state to the next is as follows: 
\begin{equation}
\label{buffer state}
    q_t^n  = \max( q_{t-1}^n + T - \Delta t ,  q_{max}^n),
\end{equation}
where $q_t^n$ is the \textit{n}th buffer occupancy level at time $t$ in seconds, $T$ is the segment length in seconds which is added to the buffer and $\Delta_t$ is the download time and, lastly, $q_{max}^n$ is the buffer size.
\subsection{Transmission Model}
Here, we provide the downlink mm-Wave communication model between the BSs and VUs. In each time step, the BSs allocate their resource blocks (RBs) among the VUs. Thus, we can express the downlink bitrate $d_t^j$ towards the \textit{i}th VU by the \textit{j}th BS in time slot \textit{t} as follows:
\vspace{-0.2cm}

\begin{equation}
d^{i,j}_t = b_{i,j}(t)\log_2(1 + \zeta_{i,j} (t)),\hspace{0.2cm}\forall j\in\mathcal{B}  , \forall i\in\mathcal{V} 
\end{equation}
where $b_{i,j}$ is the bandwidth of the assigned RB to the \textit{i}th VU by the \textit{j}th BS in time slot \textit{t} and $\zeta_{j} (t)$ is  the signal to interference plus noise ratio (SINR) from the \textit{i}th VU to the
\textit{j}th BS in time slot \textit{t}: i.e.,
\vspace{-0.2cm}

\begin{equation}
\zeta^{i,j}_t= \frac{p_{i,j} \phi_{i,j}}{ \sigma_{i,j}^2(t)}, \hspace{0.2cm} \forall j\in\mathcal{B} , \forall i\in\mathcal{V} 
\end{equation}
where $p_{i,j}$ is the transmission power between the \textit{i}th VU and the \textit{j}th BS, $\phi_{i,j}$ is the mm-Wave path loss, %In our proposed architecture, a UAV is hovering at a fixed position on the air and the link between a UAV and a VU is assumed to experience LoS mmWave communications.
$\sigma_{i,j}^2(t) = b_{i,j}(t) \textit{N}_0$ is the thermal noise power in time slot \textit{t},
The mm-Wave path loss between the \textit{i}th VU and the \textit{j}th BS can be written as:
\begin{equation}
\phi_{i,j}= \beta_1+10\beta_2\log_{10}(\kappa_{i,j})+ \xi,  \hspace{0.2cm}\forall j\in\mathcal{B}  , \forall i\in\mathcal{V} 
\end{equation}
where $\kappa_{i,j}$ is the 3-D distance between the BS and the VU;
$\xi$ is the lognormal shadowing; and $\beta_1$ and $\beta_2$ are environmental constants \cite{mmwave2018}.
\vspace{-0.2cm}
\vspace{-0.2cm}

\section{Problem Analysis}
\begin{table}[t]
    \centering
    \vspace{0.2in}
    \renewcommand{\arraystretch}{1.2}
    \setlength{\arrayrulewidth}{0.3mm}
    \setlength{\tabcolsep}{6pt}
    \caption{List of Notation.}
    \label{tab:notations}
    \begin{tabular}{|c|l|}
        \hline
        \textbf{Symbol} & \textbf{Definition} \\ 
        \hline
        $\mathcal{B}$ & Set of BSs. \\
        \hline
        $\mathcal{V}$ & Set of VUs. \\
        \hline
        $\mathcal{N_L}$ & Number of video layers. \\
        \hline
        $\mathcal{N_B}$ & Number of VU's Buffers. \\
        \hline
        $q_{t}^n$ & $\textit{n}^{th}$buffer occupancy level in seconds at time \textit{t}. \\
        \hline
        $q_{max}^n$ & $\textit{n}^{th}$buffer size in seconds. \\
        \hline
        $d^{j}_t$ & Throughput of $\textit{j}^{th}$BS at time  \textit{t}. \\
        \hline
        $i^n_t$ & Index of last downloaded segments of each layer at time  \textit{t}. \\
        \hline
        $h^j_t$ & History of Throughput values for $\textit{j}^{th}$BS at time \textit{t}. \\
        \hline
        $\mathcal{H}_0$ & Target Rebuffering time. \\
        \hline
        $\mathcal{H}_1$ & Target quality variation. \\
        \hline
    \end{tabular}
    
\end{table}

\subsection{Constrained MDP Formulation}
We model the QoE maximization problem as a discrete-time Constrained Markov Decision Process (CMDP) and analyze its mathematical formulation. A CMDP models a dynamic system with state space $\mathcal{S}$, action space $\mathcal{A}$, reward function $R : \mathcal{S} \times \mathcal{A} \rightarrow \mathbb{R}$, and transition probability function $P(s_{t+1} \mid s_t,\alpha_t)$. Our goal is to find an optimal policy that maximizes the expected discounted cumulative reward subject to the cost constraints over time. 
\subsubsection{State Set}
%The state space includes relevant system parameters that influence decision-making: $q_{t}^n $ is a vector representing the each buffer level in seconds. $d^{j}_t$ is a vector containing the throughput value of each BS (Base Station) to VU in kbps; $i_t^n$ is a vector indicating the latest downloaded segment index for each layer; and lastly, let $H$ be the number of historical throughput records that we stored for each BS. We define $h_{t}^j$ is a matrix containing the last $H$ throughput values for $j{th}$BS at time \textit{t}, which can be written as: 
%\begin{equation}
   % h^{j}_t = \bigl[d^j_{t-1}, \; d^j_{t-2}, \;\ldots,\; d^j_{t-H}\bigr],
%\end{equation}
%Where, in this paper, we set $H=5$ to help the agent to consider previous network condition to make decisions.
%$s_t = \{q_{t}^n , r^{j}_t, i_t^n, h^{j}_t\} $

The state space includes all relevant system parameters that influence decision-making: $q_{t}=[q_t^n] \in \mathbb{R}_{+}^{\mathcal{N}_B}$ is a vector representing the buffer levels in seconds, where $\mathcal{N}_B$ is the number of buffers. Let $d_{t}=[d_t^j] \in \mathbb{R}_{+}^{|\mathcal{B}|}$ be a vector containing the achievable throughput (in kbps) from each BS to the VU, $i_t=[i_t^n] \in \mathbb{Z}_{+}^{\mathcal{N}_L}$ is a vector indicating the latest downloaded segment index for each layer, where $\mathcal{N}_L$ is the number of video layers. Lastly, let $H$ be the number of historical throughput records stored for each BS. We define $h_{t} \in \mathbb{R}^{|\mathcal{B}| \times H}$ as a matrix containing the last $H$ throughput values for each BS $j$ at time \textit{t}, which can be written as:  
  \begin{equation}
      h^{j}_t = \bigl[d^j_{t-1}, \; d^j_{t-2}, \;\ldots,\; d^j_{t-H}\bigr], \quad \forall j \in \mathcal{B}
  \end{equation}
  In this work, we set $H=5$ to allow the agent to consider past network conditions when making decisions.

\subsubsection{Action Set} 
In this receiver-driven scenario, the client selects which video segment to download at each time step. We define the action as $\alpha_t \in \{0,1\}$, where $\alpha = 0$ indicates that the agent downloads a BL segment from the MBS and $\alpha = 1$ indicates that it downloads an EL segment from UAV.

\subsubsection{Reward and Cost components} 
The QoE as our main objective is influenced by three factors: 1) the video quality reward measured using the Peak Signal-to-Noise Ratio (PSNR); 2) the rebuffering cost, which captures the negative impact of stall events caused by BL buffer depletion; and 3) the smoothness cost, which captures the negative impact on perceived video quality due to quality fluctuations. Formally, we define the Quality reward as:
\begin{equation}
R_{\text{quality}}(s_t,\alpha_t) =
\begin{cases}
    {PSNR}_{BL}(k_t) , &\text{if} \ \ \ \alpha_t = 0,
    \\
    {PSNR}_{BL+EL}(k_t) , &\text{if} \ \ \ \alpha_t = 1.
\end{cases}
\end{equation}
Where $k_t$ is the segment index selected at time t. The rebuffering cost can be written as: %We define smoothness cost as. The rebuffering cost is defined as: Perceived quality reward subtracting from rebuffering or stall time cost and quality variation changes or simply smoothness cost. We consider the perceived quality reward for each video segment computed as Peak Signal-to-Noise Ratio (PSNR) which serves as a critical component in evaluating the visual fidelity of video streaming. For the rebuffering cost, we track the base buffer depletion as it is the main reason of stall time since the enhanced data could not be played independently.
\begin{equation}
C_{\text{buffer}}(s_t,\alpha_t) = {\mathbf{E}}\{\min(q_{t-1}^{n} + T - \Delta t(\alpha_t), 0)\}, \text{for} \hspace{0.1cm} n = 1,
\end{equation} 
%\jc{The text right after the equation is confusing. In the equation, we say $n=1$, but after that we say the buffer level for the BL is for $n=1.T$ (I guess you mean here from 1 to T. Please put dots in between these two symbols in that case.}
where $q_{t-1}^{n}$ denotes the BL buffer level when $n=1$. The segment length is denoted as $T$ (in seconds), and $\Delta t$ represents the download time. 
%Lastly, Smoothness cost which reflects quality fluctuations in streaming. 
%And 3) the smoothness cost which captures the negative impact on perceived video quality due to quality fluctuations. 
The smoothness cost can be modeled in different ways. Here, we define it as the changing status of the EL buffer (i.e., from empty to non-empty and vice versa):
\begin{equation}
C_{\text{smooth}}(s_t,\alpha_t) = \mathbf{E}\!\bigl[\mathbb{I}\bigl(q_{t}^{n}(\alpha_t),\,q_{t-1}^{n}\bigr)\bigr], \hspace{0.2cm} \text{for} \hspace{0.1cm} n > 1,
\end{equation}
where
\vspace{-0.4cm}
\begin{equation*}
\mathbb{I}(x,y) = 
\begin{cases}
1, & \text{if } \bigl(x = 0 ~\&~ y \neq 0\bigr)\ \text{or} \ \bigl(x \neq 0 ~\&~ y = 0),\\
0, & \text{otherwise}.
\end{cases}
\end{equation*}
In words, $\mathbb{I}{(\cdot)}$ is an indicator function that is set to 1 when there is a change in the EL buffer status (from empty to non-empty and vice versa) and 0 otherwise.
\subsubsection{Transition Function}
The transition probability function describes how the system state evolves based on the selected action 
and can be expressed as:
\begin{equation}\nonumber
\begin{aligned}
P(s_{t+1} \mid s_t,\alpha_t) 
=Pr\{q'|q, \alpha\} Pr\{d'|d\} Pr\{i'|\alpha\} Pr\{h'|h\}.
\end{aligned}
\end{equation}
Note that we cannot calculate the transition probability function because $Pr\{d'|d\}$ is unknown and depends on the network conditions.
\subsubsection{Problem Formulation}
We model our problem as a CMDP, aiming to maximize total video quality while ensuring that the expected long-term rebuffering and smoothness costs do not exceed predefined thresholds. Formally:
\begin{equation}
\begin{aligned}
\max_{\pi} \quad 
& \mathbb{E}\!\Biggl[
    \sum_{n=0}^{\infty} \gamma^{n}\, 
    R_{\text{quality}}\!\bigl(s_n,\, \pi(s_n)\bigr)
\Biggr], \\
\text{s.t.} \quad 
& \mathbb{E}\!\Biggl[
    \sum_{n=0}^{\infty} \gamma^{n}\, 
    C_{\text{buffer}}\!\bigl(s_n,\, \pi(s_n)\bigr)
\Biggr]
\,<\,{\mathcal{H}}_{0}, \\
& \mathbb{E}\!\Biggl[
    \sum_{n=0}^{\infty} \gamma^{n}\,
    C_{\text{smooth}}\!\bigl(s_n,\, \pi(s_n)\bigr)
\Biggr]
\,<\, {\mathcal{H}}_{1},
\end{aligned}
\end{equation}
where $\gamma \in (0,1)$, ${\mathcal{H}_0}$, and ${\mathcal{H}_1}$ represent the long-term target rebuffering time and the long-term target quality variation. We enforce these constraints via the Lagrange multiplier method \cite{badnava2024} which take our cost weights $\eta, \mu$ as the dual variables that adaptively penalize rebuffering and smoothness in the unconstrained reward function. Thus, the immediate QoE at each step is given by:
\begin{equation}\nonumber
r(s_t, \alpha_t) = \lambda R_{\text{quality}}(s_t,\alpha_t) - \eta C_{\text{buffer}}(s_t,\alpha_t) - \mu C_{\text{smooth}}(s_t,\alpha_t),
\end{equation}
where $(\lambda, \eta, \mu)>0$ are the video quality, rebuffering, and smoothness weights, respectively.

\subsubsection{Policy and Value Function}
The objective of our CMDP is to find a decision policy $\pi: \mathcal{S} \rightarrow \mathcal{A}$ that specifies which action to take in each state to maximize the overall QoE, which reflects the trade-off between video fidelity, rebuffering and smoothness costs. Formally, we define the optimal value function $V^{*}(s)$ under the optimal policy $\pi^*$ as the expected sum of discounted rewards when starting from state $s$:
\begin{equation}\nonumber
V^*(s_t) = \max_{\alpha_t \in \mathcal{A}} \left[ r(s_t,\alpha_t) + \gamma \sum_{s_{t+1}} P(s_{t+1} \mid s_t,\alpha_t) V^*(s_{t+1}) \right]
\end{equation}
where $\gamma \in (0,1)$ is the discount factor. Due to the large state space, the complex relation of multiple reward components, and the fact that $P(s_{t+1}\mid s_t,\alpha_t)$ is unknown, standard dynamic programming techniques are \emph{not} applicable. Instead, we rely on a \textit{deep reinforcement learning} approach in which the agent learns both the value function and policy directly through interactions with the environment. We adopt a policy gradient-based algorithm that iteratively updates the policy $\pi$ and achieves robust performance in our system.

\begin{algorithm}[t]
\caption{\emph{ASL360}: PPO with Dynamic Weight Adjustment}
\label{alg:PPO}
\begin{algorithmic}[1]
\STATE \textbf{Initialize} policy network $\pi$, value function $V$
\STATE Initialize weights and target values $\eta_{old}$, $\mu_{old}, \mathcal{H}_0, \mathcal{H}_1$ 
\FOR{each episode}
    \STATE Initialize state $s_0$
    \FOR{each time step $t$ in the episode}
        \STATE Select action $\alpha \sim \pi(\alpha | s)$
        \STATE Execute $\alpha$, observe reward $r$ and next state $s'$
        \STATE Store transition $(s, \alpha, r, s')$
        \STATE Dynamic weight Adjustment:
        $\{\eta_{new}, \mu_{new}\}$
    \ENDFOR
    \FOR{each epoch}
        \STATE Compute advantage estimate, $\hat{A}_t$
        \STATE Compute clipped objective function,
        $\mathcal{L}^{\mathrm{CLIP}}$
        \STATE Update policy and value function
    \ENDFOR
\ENDFOR
\end{algorithmic}
\end{algorithm}
\subsection{Our Proposed Method}
%Since the problem is a non-convex, non linear and mixed discrete optimization problem, it is difficult to find the optimal scheduling to maximize the QoE. Thus, we employ a DRL method to find the optimal policy. Specifically we implemented the PPO algorithm [ppo], which optimizes policy updates through a clipped objective function. The agent selects actions from a stochastic policy and updates its parameters using advantage estimation. 
Building on the CMDP formulation, we adopt the Proximal Policy Optimization (PPO) algorithm \cite{ppo} to learn the optimal policy $\pi$ that maximizes the QoE. This algorithm uses a clipped objective function that improves the stability and sample efficiency. 
At each training iteration, PPO maintains an \textit{old} policy $\pi_{{\mathrm{old}}}$ while collecting experience tuples $\{(s, \alpha, r, s')\}$. Let $\hat{\rho}_t$ be the probability ratio between the new and old policies: i.e.,
\begin{equation}
    \hat{\rho}_t={\pi(\alpha_t \mid s_t)}/{\pi_{{\mathrm{old}}}(\alpha_t \mid s_t)},
\end{equation}
\vspace{0.2in}
Now, the PPO clipped objective is given by:
\begin{equation}
\mathcal{L}^{\mathrm{CLIP}}
=\hat{\mathbb{E}}_t
\Bigl[
    \min\Bigl(\hat{\rho}_t\,\hat{A}_t,\,
             \mathrm{clip}(\hat{\rho}_t,\,1-\epsilon,\,1+\epsilon)\,\hat{A}_t
        \Bigr)
\Bigr],
\end{equation}
where $\epsilon$ is a small hyperparameter and $\hat{A}_t$ is the advantage estimate at time $t$, defined by:
\begin{equation}
    \hat{A}_t = r_t + \gamma\,V(s_{t+1}) - V(s_t),
\end{equation}
In parallel, we train value function $V(s_t)$ following standard PPO practices \cite{ppo}. By restricting policy updates within the range $[ 1-\epsilon,\,1+\epsilon ]$, PPO limits destructive changes of the policy, yielding more stable training. 
\vspace{-0.2cm}
\subsection{Dynamic Weights Adjustment}
To solve the constrained MDP formulation, we exploit a dynamic weight adjustment mechanism derived from the Lagrange multiplier method \cite{badnava2024}. At each episode, the agent observes both reward and costs and updates the cost weights $(\eta, \mu)$, so that rebuffering and smoothness remain close to their respective target rebuffering time $\mathcal{H}_0$ and the target quality variation $\mathcal{H}_1$:
 
\begin{equation}
\begin{aligned}
    \eta_{\text{new}} &= \eta_{\text{old}} + \omega_0 \left(\mathcal{H}_0 - C_{\text{buffer}}(s_t,\alpha_t)\right), \\
    \mu_{\text{new}} &= \mu_{\text{old}} + \omega_1 \left(\mathcal{H}_1 - C_{\text{smooth}}(s_t,\alpha_t)\right),
\end{aligned}
\end{equation}
where $\omega_0, \omega_1>0$ are update learning rates. 
This approach ensures that the agent aligns its optimization priorities with the streaming objectives. For instance, when the BL buffer is low, the system raises the rebuffering weight to avoid stalls. Once the buffer stabilizes, it lowers these weights, encouraging the agent to select EL segments.
\begin{figure}[t]
\centering
\includegraphics[width=0.4\textwidth]{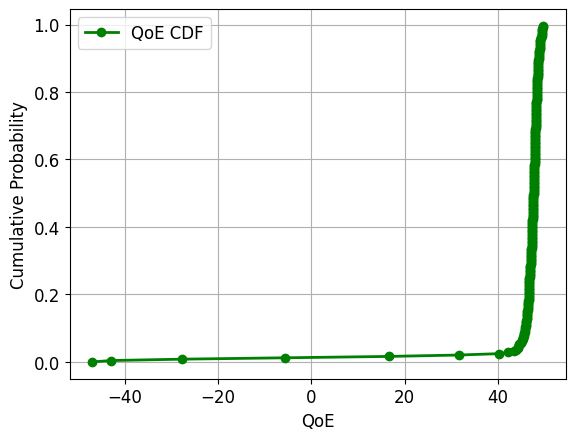}
\vspace{-0.35cm}
\caption{CDF of QoE (total reward) per episode.}
\label{cdf}
\end{figure}
\begin{figure}[t]
\centering
\includegraphics[width=0.38\textwidth]{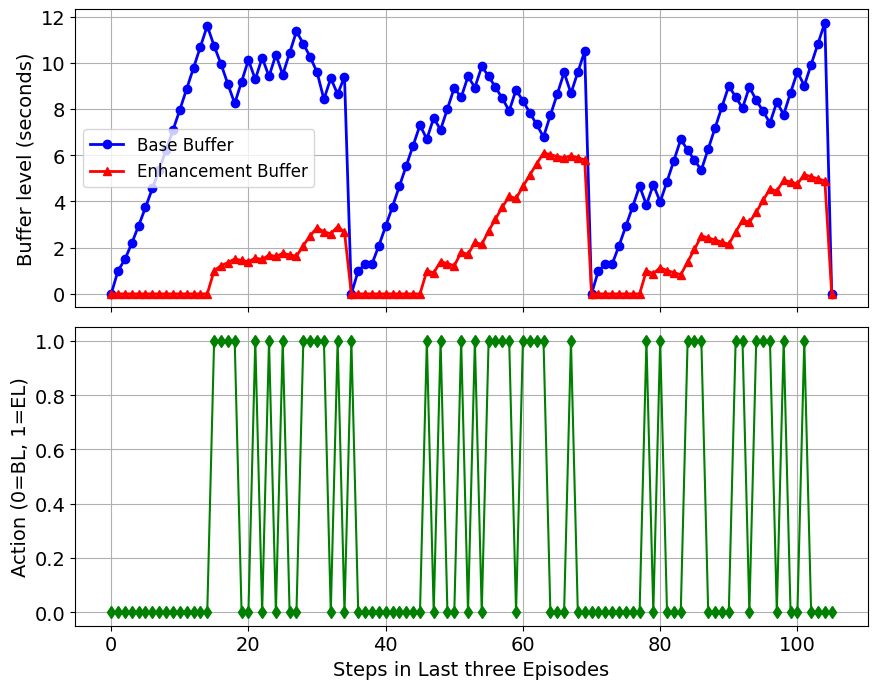}
\caption{(Top) Base and enhancement layer buffer level evolution in last 3 episodes; (bottom) Corresponding actions taken by the agent (0 = BL, 1 = EL).}
\label{buffer-evolution}
\end{figure}

\begin{figure*}[h]
\centering
\includegraphics[width=0.8\textwidth]{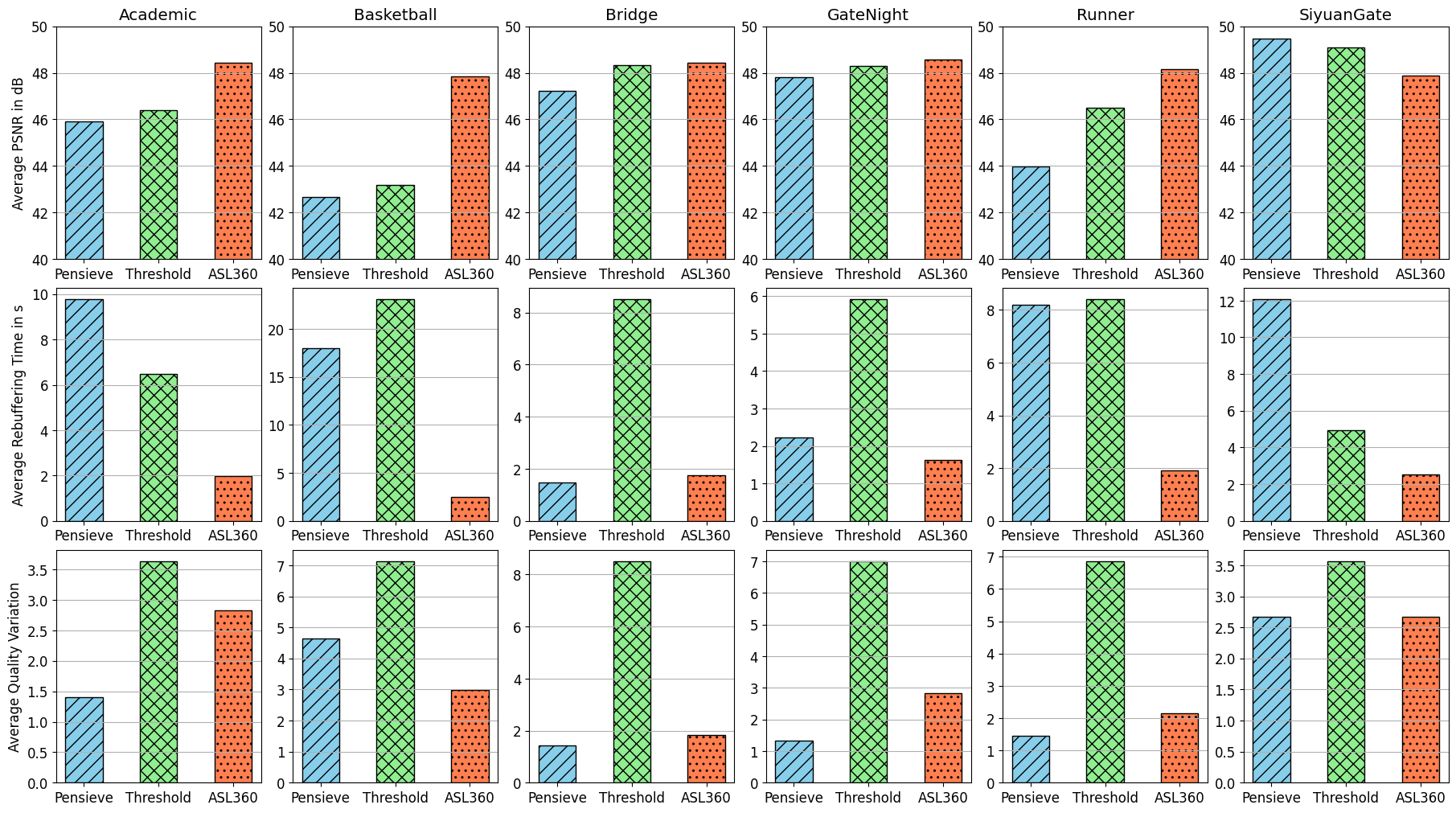}
\caption{ASL360 performance compared to the baseline methods in average quality, rebuffering time, and quality variation.}
\vspace{-3pt}
\label{compare}
\end{figure*}
\section{Simulation Results}%\label{AA}
Here, we present our results and compare them with different baselines. In our experimental setup, we consider a single VU ($|\mathcal{V}|=1$) served by two base stations: a MBS and a UAV ($|\mathcal{B}|=2$). To maintain computational and compression efficiency, we include only one EL along with the BL ($\mathcal{N_L}=2$). The VU utilizes two separate buffers for the BL and EL ($\mathcal{N_B}=2$) to store their respective video segments. The buffers have a fixed length of 36 seconds, equal to the video duration.

We utilized the SJTU 8K \tsd videos \cite{liu2017} along with the bitrate and Y-MSE distortion dataset provided for each tile-GOP in \cite{jacob2021-2}. Each video sequence has a duration of 36 seconds, comprising 1080 frames at 30 frames per second, with each GOP containing 30 frames ($T=1s$). For the BL, all tiles of each frame are encoded using the quantization parameter $QP=35$ whereas, for the EL, only tiles within the viewport are encoded using a lower $QP=15$ providing higher quality and bitrate. The viewport tiles were identified based on real head-movement traces \cite{jacob2021-2}. Moreover, instead of relying on synthetic path-loss and explicitly modeling the UAV's trajectory, we use empirical $5G$ mm-Wave throughput traces from Lumos5G \cite{Lumos5G} for both the MBS and the UAV.  
%were sampled directly at each time step during simulations. 
This approach supports realistic modeling and allows us to focus on scheduling and QoE optimization without  any mobility/channel assumptions.
For training our PPO model, we set the learning rate to 0.001, discount factor $\gamma=0.95$, clip parameter $\epsilon=0.2$ and use generalized advantage estimation (GAE) for advantage computation. The target rebuffering time $\mathcal{H}_0$ and target smoothness $\mathcal{H}_1$ are set as 0 and 1, respectively with corresponding update rates $\omega_0 = \omega_1 = 0.01$. All experiments are conducted on a server with a 13th-gen Intel CPU and an NVIDIA RTX 4090 GPU. Figure~\ref{cdf} illustrates the cumulative distribution function (CDF) of total collected QoE per episode. Our PPO-based framework, prevents any large, destructive policy updates and the policy  converges toward optimal decisions very fast across various network conditions.

Figure~\ref{buffer-evolution} illustrates the evolution of buffer levels during the last three episodes after policy convergence.
The top subplot shows the BL and EL buffer levels, while the bottom shows the agent’s action (0 for BL, 1 for EL). At first, the agent fills the BL buffer to  reduce rebuffering risk. Then, it schedules more EL segments to enhance video quality. The buffer trajectories indicate that our method successfully maintains a balance, ensuring minimal rebuffering while delivering high-quality video. This behavior arises from the flexibility introduced by the layered, tiled \tsd video structure and our adaptive weight adjustment mechanism, which allows the RL agent to optimize the cost function and maximize the overall reward. 

We evaluate our proposed method against two baselines to illustrate its performance advantages. The first, \textit{Threshold-based} method, employs a straightforward buffer-management policy: the VU keeps the BL buffer above a predefined safe threshold (we consider 10 seconds). Once this threshold is reached, the VU is permitted to download EL segments.  Such threshold-based buffer policies are common in video streaming as a practical way to prevent rebuffering and smooth playback\cite{bura2024}. 
The second baseline is \emph{Pensieve}\cite{pensieve}, a reinforcement learning-based adaptive bitrate (ABR) method originally proposed for 2D video streaming. However, Pensieve has demonstrated robust performance across diverse network conditions, making it a relevant baseline for comparison. We implemented Pensieve, without modification, allowing it to choose between low quality segments (equivalent to BL segments) and high quality segments (the entire frames are encoded at higher quality) based solely on its learned RL policy. 
Figure~\ref{compare} compares \emph{ASL360} with the baselines across six \tsd videos from our dataset in terms of average PSNR, average rebuffering time and average quality variation. Each column shows performance on a different video over several network traces, with the first row indicating the delivered average video quality (higher PSNR is better), the second row showing the rebuffering time in seconds (lower is better), and the third row reflecting the average number of quality variations (lower is better). On average, \emph{ASL360} achieves 2 dB higher PSNR than Pensieve and the Threshold-based method across videos 1-5, demonstrating its superior visual fidelity. Additionally, \emph{ASL360} shows significantly lower rebuffering time around 78-80\%, resulting in a more stable playback with fewer stalls. Finally, \emph{ASL360} maintains around 57\% lower quality variation compared to the Threshold-based method, which frequently oscillates between low and high quality. The other observation is related to the 6$^{th}$ video, \emph{ASL360} dynamically adjusts its cost weights to reduce stalls while maintaining high video quality. In contrast, Pensieve achieves only a minor improvement in quality at a much higher rebuffering cost. We attribute \emph{ASL360}'s superior performance to our layered scheduling strategy for tiled 360° videos. By delivering high-quality segments only to the user’s viewport tiles and maintaining a safe BL buffer level, \emph{ASL360} minimizes stalls more than the baselines and allows for more efficient bandwidth usage, that maximized perceived quality. 

\begin{figure}[t]
\centering
\includegraphics[width=0.4\textwidth]{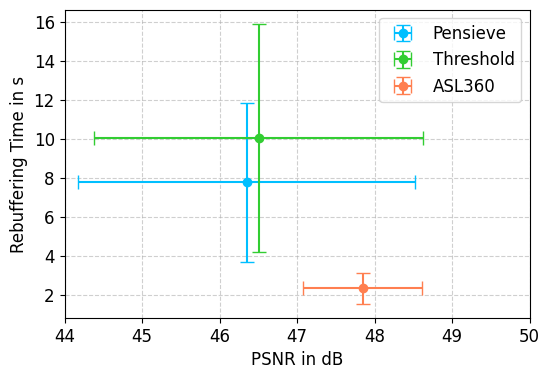}
\caption{Rebuffering time vs. quality across network traces.}
\label{rebuf-quality}
\end{figure}

\begin{figure}[t]
\centering
\includegraphics[width=0.39\textwidth]{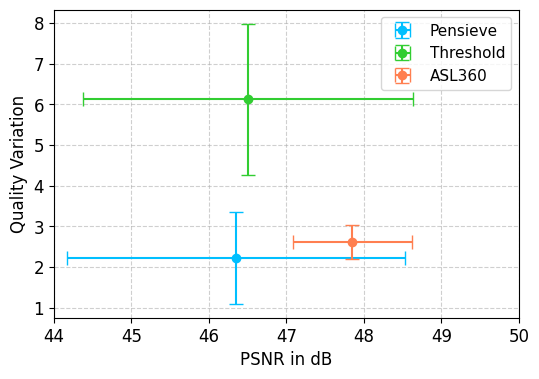}
\caption{Quality variation vs.  quality across  network traces.}
\label{smooth-quality}
\end{figure}

Figures~\ref{rebuf-quality} and~\ref{smooth-quality} show the observed range of quality against rebuffering time and quality variation across different network traces. \emph{ASL360} achieves the highest average quality and lowest average rebuffering time, while also maintaining the least variation of these metrics across network traces demonstrating a balance between quality reward and rebuffering cost. By learning to operate within the optimal scheduling range of our layered structure, \emph{ASL360} maximizes QoE across diverse scenarios.
We can also observe that, although Pensieve displays a close smoothness level performance, it does so at the cost of lower quality and higher rebuffering time.

%Consequently, these results confirm that \emph{ASL360} outperform the baselines and provides a more balanced and robust \tsd streaming experience, delivering higher quality, minimal stall events, and fewer abrupt quality changes than the baseline methods. 
\vspace{-0.1cm}
\section{Conclusion}%\label{AA}
We proposed \emph{ASL360}, an adaptive scheduling framework leveraging deep reinforcement learning for efficient on-demand streaming of layered \tsd video over UAV-assisted wireless networks. %Specifically, we considered a wireless network consisting of a macro base station (MBS) and a UAV-mounted base station (UAV), which serve multi-layered video content to VR users who employ multiple buffers to store corresponding video layers.We formulated the receiver-driven scheduling task as a MDP and employing our algorithm to optimize the reward function as the user's QoE including video fidelity, rebuffering time and quality variations. 
Through extensive simulations over realistic $5G$ mm-Wave traces and 8K video sequences,  \emph{ASL360} demonstrated a superior performance compared to the Threshold-based method and RL-based solution (\emph{Pensieve}). In particular, \emph{ASL360} improved average perceived video quality by approximately 2 dB in terms of PSNR and substantially reduced the average rebuffering time by nearly 78–80\% over baselines. Additionally, our approach reduced the quality variation by approximately 57\% compared to Threshold-based. These results highlight the capability of our layered \tsd video streaming solution. 
For future work, we plan to enhance our framework by integrating computing resources into \emph{ASL360} to jointly manage network and computational resource allocation to maximize users' QoE in immersive video services. Moreover, we aim to evaluate the scalability of \emph{ASL360} by increasing the number of VUs and investigating the optimal resource allocation algorithm for efficient \tsd video streaming. In multi-user scenarios, our multi-agent PPO will learn a joint layer-selection and resource allocation policy with QoE fairness constraints among all VUs.
\vspace{-0.6cm}
\bibliography{new_conference_101719}
\end{document}